\documentclass[preprintnumbers, prd, showpacs, floatfix,twocolumn,
preprintnumbers, letterpaper, superscriptaddress,nofootinbib]{revtex4}
\usepackage{graphicx,epsfig}
\usepackage{amsmath}
\usepackage{amssymb}
\usepackage{subfigure}
\usepackage{epstopdf}
\begin{document}

\title{Einstein-Gauss-Bonnet traversable wormholes satisfying the weak
energy condition}
\author{Mohammad Reza Mehdizadeh}
\email{mehdizadeh.mr@uk.ac.ir}
\affiliation{Research Institute for Astronomy and Astrophysics of Maragha (RIAAM), P.O.
Box 55134-441, Maragha, Iran}
\affiliation{Department of Physics, Shahid Bahonar University, P.O. Box 76175, Kerman,
Iran}
\author{Mahdi Kord Zangeneh}
\email{mkzangeneh@shirazu.ac.ir}
\affiliation{Physics Department and Biruni Observatory, College of Sciences, Shiraz
University, Shiraz 71454, Iran}
\author{Francisco S. N. Lobo}
\email{fslobo@fc.ul.pt}
\affiliation{Instituto de Astrof\'isica e Ci\^encias do Espa\c{c}o, Universidade de
Lisboa, Faculdade de Ci\^encias, Campo Grande, PT1749-016 Lisboa, Portugal}
\date{\today }

\begin{abstract}
In this paper, we explore higher-dimensional asymptotically flat wormhole
geometries in the framework of Gauss-Bonnet (GB) gravity and investigate the
effects of the GB term, by considering a specific radial-dependent redshift
function and by imposing a particular equation of state. This work is
motivated by previous assumptions that wormhole solutions were not possible
for the $k=1$ and $\alpha < 0$ case, where $k$ is the sectional curvature of
an $(n-2)$-dimensional maximally symmetric space, and $\alpha$ is the
Gauss-Bonnet coupling constant. However, we emphasize that this discussion
is purely based on a nontrivial assumption that is only valid at the
wormhole throat, and cannot be extended to the entire radial-coordinate range. 
In this work, we provide a counterexample to this claim,
and find for the first time specific solutions that satisfy the weak energy
condition throughout the entire spacetime, for $k=1$ and $\alpha < 0$. 
In addition to this, we also present other wormhole solutions which alleviate
the violation of the WEC in the vicinity of the wormhole throat.
\end{abstract}

\maketitle

\section{Introduction}

Wormholes are nontrivial throatlike geometrical structures which connect two
parallel universes or distant parts of the same universe. In 1988, Morris
and Thorne introduced a family of traversable wormholes \cite{mt}, where the
fundamental ingredient is the flaring-out condition of the wormhole throat.
This latter condition, in framework of general relativity (GR), entails the
violation of the null energy condition (NEC), which states that $%
T_{\mu\nu}k^\mu k^\nu \geq 0$, where $T_{\mu\nu}$ is the energy-momentum
tensor and $k^\mu$ is \textit{any} null vector. Matter that violates the NEC
is denoted by \textit{exotic matter} \cite{snm,hoc}. Due to the problematic
issue of the violations of the energy conditions \cite{Barcelo:2002bv},
several avenues of research have been explored in order to minimize the
usage of exotic matter \cite{Visser:2003yf}. For instance, it was shown that
dynamical spherically symmetric wormholes can satisfy the energy conditions 
\cite{karsa} and the averaged energy conditions over timelike or null
geodesics for a period of time \cite{tip}. 
Another interesting construction are the thin-shell
wormholes, where the exotic matter is restricted to the throat, and
therefore minimize its usage \cite{kis}.

It was also found that
higher-dimensional cosmological wormholes \cite{kordloboriazi} and wormholes
in modified gravity, involving higher order curvature invariants, can
satisfy the energy conditions \cite{modgravWH,Harko:2013yb,oli,anc,dzh}, at
least at the throat. In fact, in modified gravity, it was shown that matter
threading the wormhole throat can be imposed to satisfy all of the energy
conditions, and it is the higher order curvature terms, which may be
interpreted as a gravitational fluid, that support these nonstandard
wormhole geometries. 
Thus, one is motivated in exploring wormhole geometries in
higher-dimensional theories, due to the fact that these alleviate the
violation of the energy conditions, at least at the throat. Of particular
interest are the $n$-dimensional Lorentzian wormhole
geometries \cite{deh} that were explored in Lovelock gravity \cite{lov}, 
which is the most general theory of gravitation in $n$ dimensions. 
In contrast to Einstein gravity, it
was found that the wormhole throat radius has a lower limit that depends on
the Lovelock coefficients, the dimensionality of the spacetime and the shape
function. In addition to this, it was shown that the higher order Lovelock
terms with negative coupling constants enlarge the region of normal matter
near the throat.

Lorentzian wormhole solutions were also investigated in the context of the $%
n $-dimensional Einstein-Gauss-Bonnet (GB) theory of gravitation, which is
the second order Lovelock gravity \cite{bha}. These wormholes were found to
have features depending on the dimensionality of the spacetime, $n$, and the
GB coupling constant, $\alpha$. It was shown that in a large number of
cases, the wormhole throat radius is constrained by $n$ and $\alpha$. The
possibility of obtaining solutions with normal and exotic matter limited to
the vicinity of the throat was also explored. Similar to the situation in
GR, the violation of the weak energy condition (WEC) persists for $\alpha>0$. For $%
\alpha<0$, this condition may or may not be violated depending on the nature
of an inequality involving $|\alpha|$, $n$, the radius $r$, and the wormhole
shape function. Dynamic wormhole solutions in the framework of Lovelock
gravity with compact extra dimensions were also analysed \cite{meh}. It was
shown that as the wormhole inflates with the three-dimensional space, the
extra dimensions deflate to very small, yet nonvanishing scales. In addition
to this, it was also shown that the WEC holds for
certain ranges of the free parameters of the theory. Further higher
dimensional wormhole solutions have been explored, and we refer the reader
to \cite{dot,kan} for more details.

A thorough analysis of the properties of $n$-dimensional static wormhole
solutions was investigated in Einstein-Gauss-Bonnet gravity with or without
a cosmological constant. The analysis in \cite{bha} was generalized in \cite%
{mae} by assuming that the spacetime possessed symmetries corresponding to
the isometries of an $(n-2)$-dimensional maximally symmetric space with the
sectional curvature $k=\pm 1,0$. The metric was assumed to be
least $C^2$ and the $(n-2)$-dimensional maximally symmetric subspace
to be compact. The solutions were classified into general relativistic (GR) and non-GR branches, 
respectively, depending on the existence or absence of the general
relativistic limit $\alpha \rightarrow 0$. The authors showed the that 
branch surface in the GR branch coincides with the wormhole throat respecting the dominant 
energy condition (DEC), otherwise the NEC is violated. On the other hand, in the non-GR branch 
for $k\alpha \geq 0$, it was shown that there is no wormhole solution. In addition to this, it was also shown in the non-GR branch with $k\alpha \leq 0$ and $\Lambda 
\leq 0$, for the matter field with zero tangential pressure, that the DEC holds at the 
wormhole throat if the throat radius satisfies a specific inequality. 
Furthermore, explicit wormhole solutions respecting the energy conditions in the whole spacetime were obtained in the vacuum and dust cases with $k=-1$ and $\alpha > 0$.

We emphasize that despite the fact that wormhole solutions satisfying the
energy conditions for the specific case of $k=-1$ and $\alpha > 0$, as
mentioned above, no solutions were found for the $k=1$ and $\alpha < 0$,
which was the case extensively explored in Ref. \cite{bha}. In the latter,
the authors claimed that wormholes solutions were not possible for $k=1$ and $%
\alpha < 0$. However, as also mentioned in \cite{mae}, this discussion is
purely based on a nontrivial assumption, which seems to be valid only at the
wormhole throat, and cannot be extended throughout the entire range of the
radial coordinate. Although, no counterexample was provided in \cite{mae},
here we provide for the first time a specific solution for wormholes that satisfy
the WEC for the specific case of $k=1$ and $\alpha < 0$.

In addition to this, in all of the above GB wormhole studies, the redshift
function is considered to be zero. Here, we relax this assumption and
consider a specific radial-dependent choice for the redshift function, which
tends to zero at spatial infinity. Indeed, since the GB term has low effects
on regions far from throat, one can expect that a non-constant redshift
function may contribute to solutions satisfying the energy conditions. Thus,
in this paper, we discuss higher dimensional wormhole solutions in the
framework of GB gravity and investigate the effects of the GB term, with the
presence of $r$-dependent redshift functions and considering a specific
equation of state, on the satisfaction of the WEC.

The paper is organized as follows: In Sec. \ref{Sol}, we give a brief review
of the field equations of GB gravity, and introduce an equation of state in
order to solve the field equations. In Sec. \ref{exact}, several
wormhole solutions are presented, more specifically, by considering
particular choices for the parameters of the theory. In Sec. \ref{conc}, we
summarize and discuss our results.

\section{Action and field equations}
\label{Sol}

\subsection{Action}

The action in the framework of GB theory, in the presence of a cosmological
constant, is given by 
\begin{equation}
I_{G}=\int d^{n}x\sqrt{-g}\left[R-2\Lambda +\alpha _{2}\mathcal{L}_{GB}%
\right],  \label{action}
\end{equation}
where $n$ is the dimension of the space-time; $R$ and $\Lambda$ are the $n$%
-dimensional Ricci scalar and the cosmological constant, respectively; $%
\alpha _{2}$ is the Gauss-Bonnet (GB) coefficient, and the GB term $\mathcal{%
L}_{GB}$ is given by 
\begin{equation}
\mathcal{L}_{GB}=R^{2}-4R_{\mu \nu }R^{\mu \nu }+R_{\mu \nu \rho \sigma
}R^{\mu \nu \rho \sigma }.
\end{equation}
In Lovelock theory, for each Euler density of order $\bar{k}$ in $n$
dimension space-time, only terms with $\bar{k}<n$ exist in the equations of
motion \cite{shek}. Therefore, the solutions of the Einstein-Gauss-Bonnet
theory are in $n\geq 5$ dimensions. Note that the action (\ref{action}) is
derived in the low energy limit of string theory \cite{myg}.

Now, varying the action (\ref{action}) with respect to metric, one obtains
the field equations 
\begin{equation}  \label{EGBfieldeq}
G_{\mu \nu }+\alpha _{2}\mathcal{G}_{\mu \nu }=T_{\mu \nu },
\end{equation}%
where $T_{\mu \nu }$ is the energy-momentum (EM) tensor, $G_{\mu \nu}$ is
Einstein tensor and $\mathcal{G}_{\mu \nu }$ is the GB tensor, given by 
\begin{eqnarray}
\mathcal{G}_{\mu \nu } &=&2(-R_{\mu \sigma \kappa \tau }R_{\phantom{\kappa
\tau \sigma}{\nu}}^{\kappa \tau \sigma }-2R_{\mu \rho \nu \sigma }R^{\rho
\sigma }  \notag \\
&&-2R_{\mu \sigma }R_{\phantom{\sigma}\nu }^{\sigma }+RR_{\mu \nu })-\frac{1%
}{2}\mathcal{L}_{GB}g_{\mu \nu }.
\end{eqnarray}
We use a unit system with $8\pi G_n=1$, where $G_n$ is the $n$-dimensional
gravitational constant.

In this work, we consider the $n$-dimensional spacetime, by replacing the
two-sphere \cite{mt} with a $(n-2)$-sphere, given by the following line
element 
\begin{equation}
ds^{2}=-e^{2\phi (r)}dt^{2}+ \frac{dr^{2}}{1-b(r)/r} +r^{2}d\Omega
_{n-2}^{2}\,,
\end{equation}
where $d\Omega_{n-2}^{2}$ is the metric on the surface of the $(n-2)$%
-sphere. $\phi (r)$ is denoted the redshift function as it is related to
the gravitational redshift; and $b(r)$ is denoted the shape function because it
determines the shape of the wormhole, as can be shown by embedding diagrams 
\cite{mt}. The radial coordinate $r$ is non-monotonic in that it decreases
from $+\infty$ to a minimum value $r_0$, which represents the throat of the
wormhole, and then increases to $+\infty$. The shape function at the throat
is defined as $b(r_0)=r_{0} $. Note that $\phi (r)$ should be finite
everywhere in order to avoid the presence of an event horizon \cite{mt}. $%
b(r)$ should satisfy the flaring-out condition, i.e., $rb^{\prime }-b<0$, so
that at the throat we verify the condition $b^{\prime }(r_0) < 1$. The
condition $1-b(r)/r \geq 0$ is also imposed. Note that although the metric
coefficient $g_{rr}$ becomes divergent at the throat, signalling a
coordinate singularity, the proper radial distance $l(r)= \pm \int_{r_0}^r
dr/(1-b/r)^{1/2}$ is required to be finite everywhere. Thus, the proper
distance decreases from $l = +\infty$, in the upper universe, to $l = 0$ at
the throat, and then from zero to $-\infty$ in the lower universe.

\subsection{Field equations}

The EM tensor is given by $T_{\nu }^{\mu }$ $=$ $\mathrm{diag}$[$-\rho $($r$%
, $t$), $p_{r}$($r$, $t$), $p_{t}$($r$, $t $), $p_{t}$($r$, $t$), $...$],
where $\rho (r)$ is the energy density and $p_{r}(r)$ and $p_{t}(r)$ are the
radial and transverse pressures, respectively. Thus, the gravitational field
equation (\ref{EGBfieldeq}) provides the following relations 
\begin{widetext}
\begin{eqnarray}
\rho (r) =\frac{(n-2)}{2r^{2}}\left\{-\left( 1+\frac{2\alpha b}{r^{3}}%
\right) \frac{(b-rb^{\prime })}{r}  +\frac{b}{r}\left[ (n-3)+(n-5)\frac{\alpha b}{r^{3}}\right] \right\},
\label{rho}
\end{eqnarray}%
\begin{eqnarray}
p_{r}(r) =\frac{(n-2)}{2r}\left\{2\left( 1-\frac{b}{r}\right) \left( 1+%
\frac{2\alpha b}{r^{3}}\right) \phi^{\prime }  -\frac{b}{r^{2}}\left[ (n-3)+(n-5)\frac{\alpha b}{r^{3}}\right] \right\},
\label{pr}
\end{eqnarray}%
\begin{eqnarray}
p_{t}(r) &=&\left( 1-\frac{b}{r}\right) \left( 1+\frac{2\alpha b}{r^{3}}%
\right) \left[ \phi''+{\phi'}^{2}+\frac{%
(b-rb')\phi'}{2r(r-b)}\right]  +\left( 1-\frac{b}{r}\right) \left( \frac{\phi ^{{\prime }}}{r}+\frac{%
b-b'r}{2r^{2}(r-b)}\right)\left[ (n-3)+(n-5)\frac{2\alpha b}{r^{3}}\right] 
\notag \\
&& -\frac{b}{2r^{3}}\left[ \left( n-3\right) \left( n-4\right) +\left(
n-5\right) \left( n-6\right) \frac{\alpha b}{r^{3}}\right] -\frac{2\phi'\alpha }{r^{4}}\left( 1-\frac{b}{r}\right)
\left( b-b'r\right) \left( n-5\right) ,\label{pt}
\end{eqnarray}%
\end{widetext}
where the prime denotes a derivative with respect to the radial coordinate $%
r $. We define $\alpha =(n-4)(n-3)\alpha _{2}$ for notational convenience.
We provide below several strategies for solving the field equations.

\subsection{Stategy of solving the field equations}
\label{sragy}

We now have three equations, namely, the field equations (\ref{rho})-(\ref%
{pt}), with the following five unknown functions $\rho (r)$, $p_{r}(r)$, $%
p_{t}(r)$, $b(r)$ and $\phi (r)$. Therefore, in order to determine the
wormhole geometry, one can adopt several strategies. For instance, one can
apply restrictions on $b(r)$ and $\phi (r)$ or on the EM tensor components.
It is also common to use a specific equation of state (EOS) relating the EM
tensor components, such as, specific equations of state responsible for the
present accelerated expansion of the Universe \cite{lob2} and the traceless
EM tensor equations of state \cite{kar2}, amongst others.

In this work, we use an equation of state (EOS) of the form \cite{APT} 
\begin{equation}
\rho =\omega \left[ p_{r}+(n-2)p_{t}\right] .  \label{sta}
\end{equation}%
Using Eq. (\ref{sta}), the trace of the EM tensor can be written as $T=-\rho
+p_{r}+(n-2)p_{t}=\rho \,(1-\omega )/\omega $. This EOS will be particularly
useful, as for $\omega=1$, it reduces to a traceless EOS, $T=0$, which is
usually associated with the Casimir effect, and that will be extensively
explored in the solutions presented below.

Now, substituting $\rho $, $p_{r}$ and $p_{t}$ in the EOS, one obtains the
following differential equation
\begin{eqnarray}
b^{\prime }(r) &=&\Big\{2r^{2}\omega (n-2)(r-b)(r^{3}+2\alpha b)\left( \phi
^{\prime 2}+\phi ^{\prime \prime }\right)  \notag \\
&&+r\omega \phi ^{\prime }\eta -\xi \Big\}/\zeta ,  \label{bprim}
\end{eqnarray}
with the following definitions 
\begin{equation*}
\eta =(n-2)\left\{ 2r^{4}(n-2)-rb\left[ r^{2}(2n-5)-4\alpha
\right] -2\alpha b^{2}\right\} ,
\end{equation*}%
\begin{eqnarray*}
\xi  &=&r^{3}b(n-2)(n-4)[(n-3)\omega +1] \\
&&+\alpha b^{2}(n-2)(n-7)[(n-5)\omega +1],
\end{eqnarray*}%
and
\begin{eqnarray*}
\zeta  =r^{4}(n-2)[(n-3)\omega +1]+2\alpha rb(n-2)[\omega (n-5)+1] \\
-(n-2)\omega r^{2}\phi ^{\prime }\left\{ (2\alpha b(-2n+9)+r\left[ 4\alpha
(n-5)-r^{2}\right] \right\} .
\end{eqnarray*}

With the EOS given by Eq. (\ref{sta}) in hand, an additional restriction is
necessary in order to close the system and solve the field equations. For
this purpose, we choose an asymptotically flat redshift function given by 
\begin{equation}
e^{2\phi (r)}=\phi _{0}+\phi _{1}\left( \frac{r_{0}}{r}\right) ^{m},
\label{phii}
\end{equation}%
where $\phi _{0}$ and $\phi _{1}$ are dimensionless constants and $m$ is a
positive constant. Note that choosing $\phi _{1}=0$, Eq. (\ref{phii})
reduces to the well-known zero tidal force case \cite{mt}. As mentioned
above, we are interested in analysing solutions that are asymptotically
flat, i.e., $b(r)/r\rightarrow 0$ and $\phi (r)\rightarrow 0$ as $%
r\rightarrow \infty $.

\section{Wormhole solutions}
\label{exact}

It is well-known that static traversable wormholes in four dimensions
violate the energy conditions at or near the wormhole throat in GR \cite%
{snm,mt,hoc}. Theses violations are derived from the flaring-out condition
of the wormhole throat. On the other hand, the energy conditions can
be satisfied in the vicinity of static wormhole throats in higher
dimensional alternative theories \cite{bha,deh} and the whole space-time in
the case of higher order curvature terms \cite{Harko:2013yb}, and dynamic
wormholes \cite{meh}. In the context of the local energy conditions, we
examine the weak energy condition (WEC), i.e., $T_{\mu \nu }U^{\mu }U^{\nu
}\geq 0$ where $U^{\mu }$ is a timelike vector. For a diagonal EM tensor,
the WEC implies $\rho \geq 0$, $\rho +p_{r}\geq 0\ $and $\rho +p_{t}\geq 0$,
simultaneously. Note that the last two inequalities are defined as the NEC.

Using Eqs. (\ref{rho})-(\ref{pt}), one finds the following
relationships 
\begin{equation}
\rho +p_{r}=-\frac{(n-2)}{2r^{2}}\left[ \frac{(b-rb^{\prime )}}{r}+2\phi
^{\prime }\left( b-r\right) \right] \left( 1+\frac{2\alpha b}{r^{3}}\right) ,
\label{EGBNEC}
\end{equation}%
\begin{widetext}
\begin{eqnarray}
\rho +p_{t}&=&-\frac{\left( b-rb^{\prime }\right) }{2r^{3}}\left( 1+\frac{
6\alpha b}{r^{3}}\right)  +\frac{b}{r^{3}}\left[ (n-3)+(n-5)\frac{2\alpha b}{r^{3}}\right]
+\phi ^{\prime }\bigg[\frac{b-rb^{\prime }}{2r^{2}}\left( 1-\frac{2\alpha b
}{r^{3}}(9-2n)\right) \notag \\
&& -\frac{b}{r^{2}}\left((n-3)+\frac{2\alpha b}{r^{3}}(n-5)\right)
+\frac{1}{r}\bigg((n-3)+\frac{2\alpha b^{\prime }(n-5)}{r^{2}}\bigg)\bigg]
+\left( 1-\frac{b}{r}\right) \left( 1+\frac{2\alpha b}{r^{3}}\right)
\left({\phi ^{\prime }}^{2}+\phi ^{\prime \prime }\right) \,,
\end{eqnarray}
\end{widetext}
respectively. One can easily show that for $\alpha =0$ and $\phi ^{\prime }=0 $ the NEC,
and consequently the WEC, are violated, due to the flaring-out condition.

Note that at the throat, one verifies 
\begin{eqnarray}
\left( \rho + p_r \right)\big|_{r=r_0}=- \frac{n-2}{2r_0^2}\left(
1-b^{\prime }_0 \right)\left( 1+\frac{2\alpha}{r_0^2} \right) \,.
\end{eqnarray}
Taking into account the condition $b_0^{\prime }<1$, and for $\alpha > 0 $,
one verifies the general condition $\left( \rho + p_r \right)\big|_{r=r_0}<0$%
. For $\alpha <0$, the NEC at the throat is also violated for the range $r_0>%
\sqrt{2|\alpha|}$. In order to impose $\left( \rho + p_r \right)\big|%
_{r=r_0}>0$, one needs to consider $\alpha < 0$, and the condition $r_0<%
\sqrt{2|\alpha|}$, which proves that one may have wormholes in GB gravity
satisfying the WEC at the throat.

\subsection{Solutions violating the WEC}

\subsubsection{Einstein Gravity}

In this section, we search for exact solutions in higher-order Einstein
gravity ($\alpha =0$) imposing $m=(n-3)$ in Eq. (\ref{phii}). For this
specific case, Eq. (\ref{bprim}) provides the following solution 
\begin{equation}
b(r)=\frac{2c_{0}\phi _{0}\left[\phi _{0}+\phi _{1}\big(\frac{r_0}{r}\big)%
^{m}\right]\left( 1+\omega m\right) +\phi _{1}^{2}r_0^{m}\big(\frac{r_0}{r}%
\big)^{m}\omega m}{\phi _{0}\left\{2\phi _{0}r^{(m-1)}\left[1+\omega m\right]%
+\frac{\phi _{1}r_0^{m}}{r}(2+\omega m)\right\} },  \label{ein1}
\end{equation}%
where $c_{0}$ is a constant of integration, and can be determined using the
condition $b(r_{0})=r_{0}$. It is clear that the solutions are
asymptotically flat, i.e., $b(r)/r\rightarrow 0$ as $r\rightarrow +\infty $.
Note that from Eq. (\ref{ein1}) for $\omega =0$ and $\phi _{0}=\phi _{1}$,
one obtains $b(r)=\frac{c_{0}}{r^{(n-4)}}$ which presents the Schwarzschild
geometry. One also verifies that by choosing $\phi _{1}=0$, the wormhole
solution of Ref. \cite{vis1} is obtained.

Choosing $\omega =1$ in Eq. (\ref{sta}) leads to a traceless EM tensor
solution. In this case, one can obtain wormhole solutions with suitable
constants $\phi _{0}$ and $\phi _{1}$ in order to avoid an event horizon.
For instance, in four dimensions Eq. (\ref{ein1}) reduces to 
\begin{equation}
b(r)=\frac{4c_{0}\phi _{0}(\phi _{0}\frac{r}{r_{0}}+\phi _{1})+{\phi
_{1}r_{0}}}{\phi _{0}(4\phi _{0}\frac{r}{r_{0}}+3\phi _{1})},
\end{equation}%
where $c_{0}$ is given by 
\begin{equation}
c_{0}=\frac{r_{0}(4\phi _{0}-\phi _{1})}{4\phi _{0}},
\end{equation}%
with $\phi _{1}+\phi _{0}>0$, to avoid the presence of an event horizon.
Using the field equations (\ref{rho})-(\ref{pt}) one obtains 
\begin{equation}
\rho =-\frac{4\phi _{1}r_{0}(c_{0}\phi _{0}+\phi _{1}r_{0})}{r^{2}(4\phi
_{0}r+3\phi _{1}r_{0})^{2}},
\end{equation}%
\begin{equation}
\rho +p_{r}=-\frac{16(r\phi _{0}+\phi _{1}r_{0})(c_{0}\phi _{0}+\phi
_{1}r_{0})}{r^{2}(4r\phi _{0}+3\phi _{1}r_{0})^{2}},
\end{equation}%
\begin{equation}
\rho +p_{t}=\frac{8\phi _{0}(c_{0}\phi _{0}+\phi _{1}r_{0})}{r(4\phi
_{0}r+3\phi _{1}r_{0})^{2}}.
\end{equation}%
It is clear that the conditions $\phi _{0}+\phi _{1}>0$ and $\frac{-4\phi
_{1}r_{0}(c_{0}\phi _{0}+\phi _{1}r_{0})}{(4\phi _{0}r_{0}+3\phi
_{1}r_{0})^{2}}<1$ (which is imposed by $b^{\prime }(r_{0})<1$) lead to the
violation of WEC.

\subsubsection{$\protect\phi _{1}=0$}

Since solving the differential equation (\ref{bprim}) is too complicated, in
general, we will consider restrictions on the redshift function. For
instance, consider a constant redshift function, i.e., $\phi _{1}=0$ in Eq. (%
\ref{phii}). Applying these simplified choices, the shape function is given
by 
\begin{equation}
b(r)=\frac{\left[ -\bar{\omega} \pm \sqrt{\bar{\omega} ^{2}+4c_{1}\alpha
\left(\bar{\omega}-2\omega \right)\big(\frac{r}{r_0}\big)^{1-n}}\right] r^{3}%
}{2\alpha \left(\bar{\omega}-2\omega \right)},  \label{so3}
\end{equation}%
where we have defined $\bar{\omega} = \omega (n-3)+1 $ for notational
simplicity, and the constant $c_1$ is given by 
\begin{equation}
c_{1}=\left[r_{0}^{2} \bar{\omega} +\alpha \left( \bar{\omega} -2 \omega
\right) \right] r_{0}^{-4}.
\end{equation}

In order to study the behavior of this solution at infinity, we consider the
approximation 
\begin{eqnarray}
1-\frac{b(r)}{r} &\simeq &1+\frac{\left(\bar{\omega} \mp \sqrt{ \bar{\omega}
^{2}}\right) r^{2}}{2\alpha (\bar{\omega} -2\omega)} +O\left(\frac{1}{r^{n-3}%
}\right)\text{.}  \label{limitinf}
\end{eqnarray}%
There are two classes of wormhole solutions corresponding to the two signs
that appear in Eq. ( \ref{limitinf}). It is obvious that these solutions are
asymptotically flat if we choose suitable signs in the equation above,
namely, $\bar{\omega} >0$ for the positive sign, and $\bar{\omega} <0$ for
the negative sign We denote these solutions $b_{+}$ and $b_{-}$,
respectively.

Now, in order to check the WEC, we first investigate the behavior of $\rho
(r)$ for large $r$, which is given by the following approximation 
\begin{equation}
\rho (r)\simeq -\frac{(n-1)(n-2)\alpha \omega c_{1}^{2}}{\bar{\omega} }
\left( \frac{r_0}{r} \right)^{2(n-1)} +O\left(\frac{1}{r^{3(n-1)}}\right).
\end{equation}
Note that $\rho (r)$ tends to zero as $r$ increases to infinity. Since $\rho
(r)$ has no real positive root, in order to find its sign, it is sufficient
to investigate the sign at infinity. In the case of the $b_{+}$ ($b_{-}$)
solution which corresponds to $\bar{\omega} >0$ ($\bar{\omega} <0$), in
order to satisfy the WEC, one finds that $\alpha \omega <0$ ($\alpha \omega
>0$).

Let us now obtain $\rho +p_{r} $ and $\rho +p_{t}$ for large $r$: 
\begin{equation}
\rho +p_{r}\simeq -\frac{c_{1}(n-2)}{\bar{\omega} } \left( \frac{r_0}{r}
\right)^{n-1} +O\left( \frac{1}{r^{2(n-1)}}\right),
\end{equation}%
and 
\begin{equation}
\rho +p_{t}\simeq \frac{c_{1}}{\bar{\omega} } \left( \frac{r_0}{r}
\right)^{n-1} +O\left(\frac{ 1}{r^{2(n-1)}}\right).
\end{equation}
It is clear that both $\rho +p_{r}$ and $\rho +p_{t}$ tend to zero as $r$
tends to infinity, with opposite signs. Therefore, in the large $r$ limit,
one of $\rho +p_{r}$ or $\rho +p_{t} $ is negative and consequently the WEC
is violated. However, we show that one can choose suitable values for the
constant parameters in order to have normal matter in the vicinity of the
throat.

In the following analysis, we consider the traceless EM tensor, $T=0$. This
is usually associated to the Casimir effect, which violates all the energy
conditions. Thus, in order to investigate the traceless EM tensor case, we
impose $\omega =1$. In this case, $b_{+}$ reduces to 
\begin{equation}
b_{+}(r)=\frac{\left[ 2-n+\sqrt{(n-2)^{2}+4c_{1}\alpha (n-4)\big(\frac{r}{%
r_{0}}\big)^{1-n}}\right] r^{3}}{2\alpha (n-4)}.  \label{so2}
\end{equation}

The behavior of $b_{+}$ in the large $r$ limit is given by 
\begin{equation}
b_{+}(r)\simeq \frac{c_{1}r_{0}^{(n-1)}}{(n-2)r^{n-4}},  \label{bpinf}
\end{equation}%
which guarantees the asymptotic flatness of the solution. In order to check
the flaring-out condition at the throat, i.e., ${b}^{\prime }{(r_{0})}<1$,
one obtains from Eq. (\ref{bprim}) that 
\begin{equation}
{b}^{\prime }{(r_{0})}=-\frac{(n-4)[r_{0}^{2}(n-2)+\alpha (n-7)]}{%
[r_{0}^{2}(n-2)+2\alpha (n-4)]}.
\end{equation}%
Note that in Einstein gravity, where $\alpha =0$, the condition ${b}^{\prime
}{(r_{0})}<1$ is satisfied independent of $r_{0}$. In the case of $\alpha >0$%
, the flaring-out condition is also satisfied, but for $\alpha <0 $ it
should be carefully checked.

\begin{figure*}[tbp]
\begin{center}
\subfigure[~$\alpha=-1$]{
    \label{k0a}
    \includegraphics[width=0.48\textwidth,height=0.3\textheight]{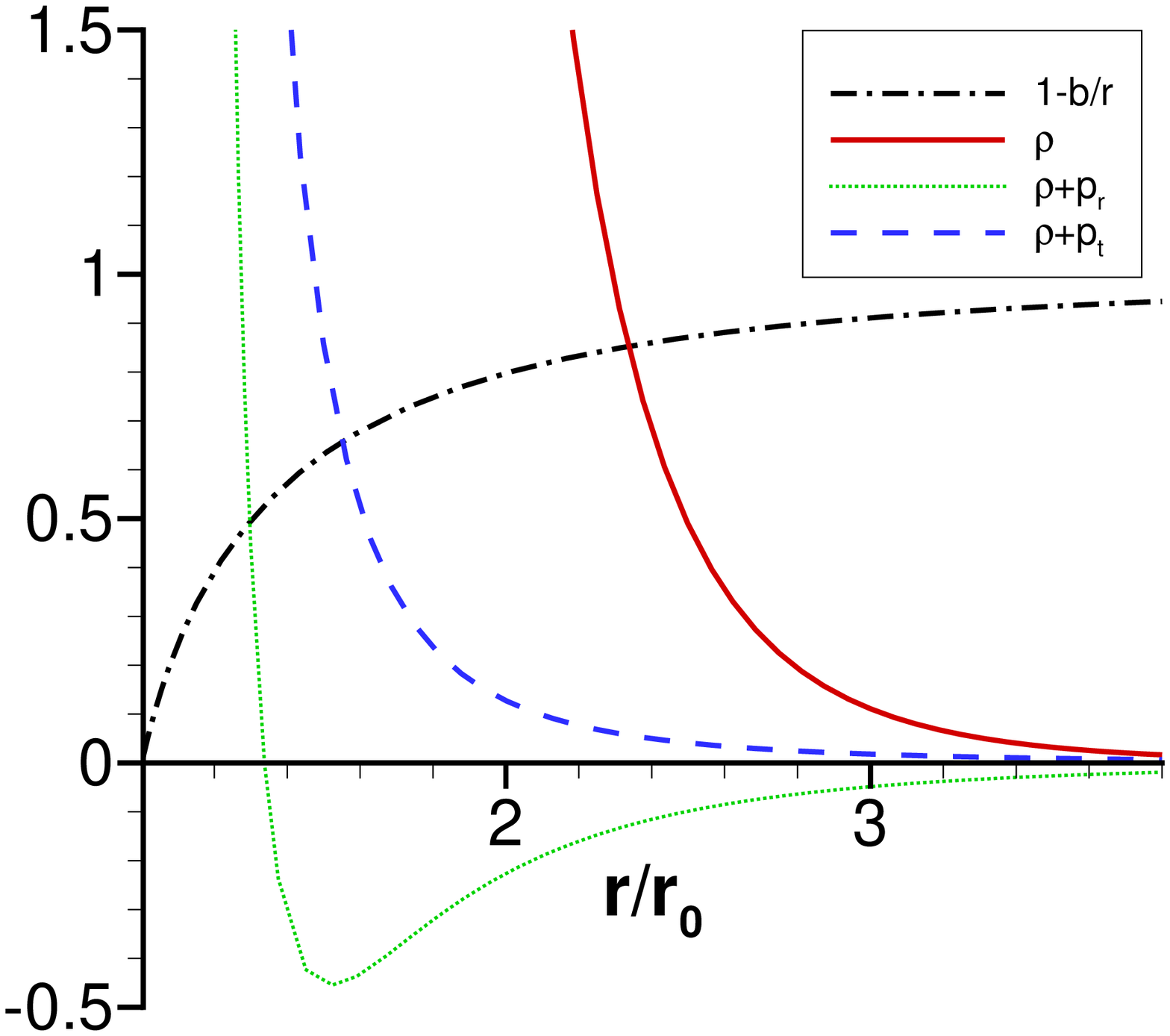}} 
\text{\hspace{0cm}} 
\subfigure[~$\alpha=1$]{
   \label{k1a}
   \includegraphics[width=0.48\textwidth,height=0.3\textheight]{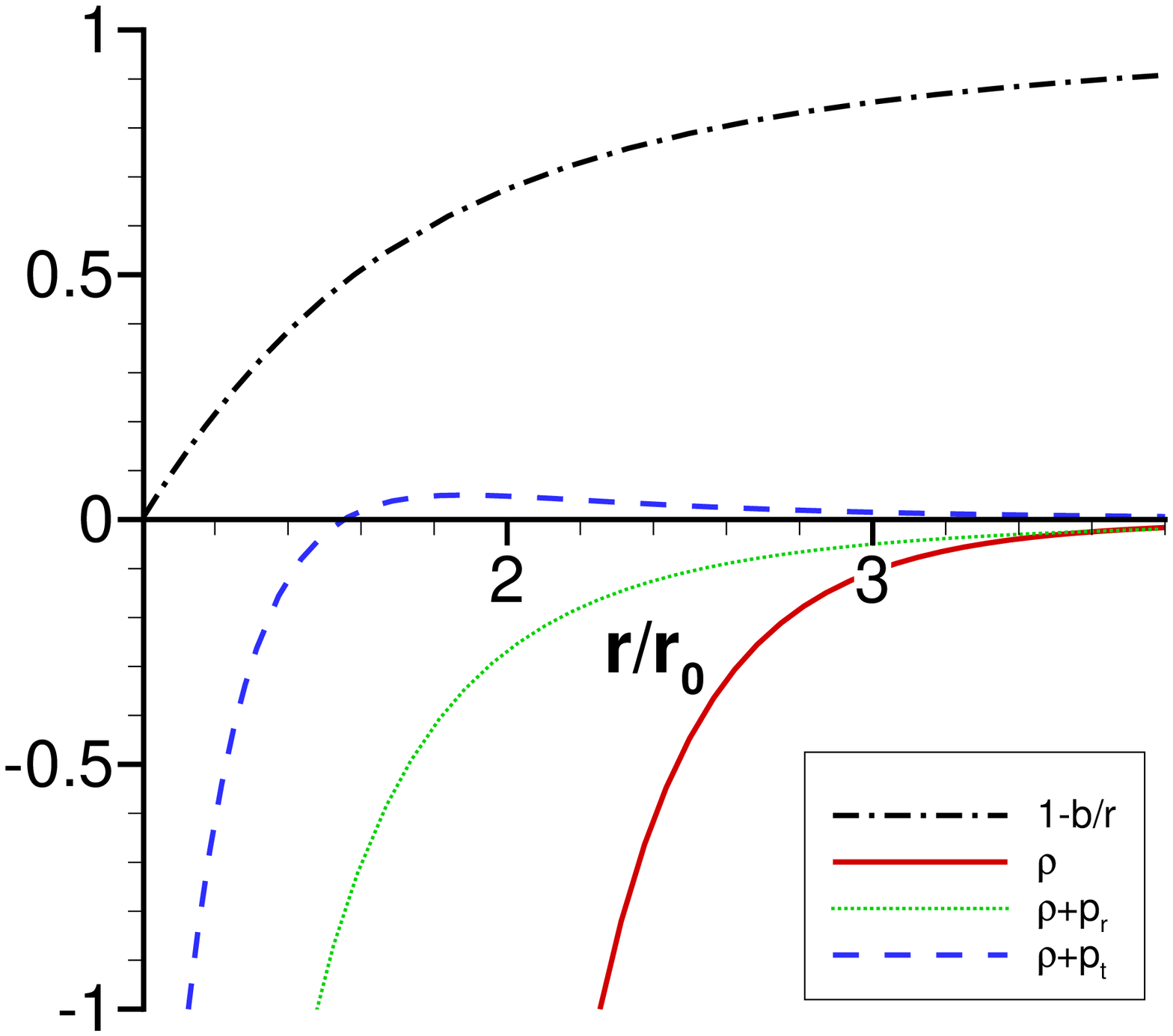}}
\end{center}
\caption{The specific case of a constant redshift function, $\protect\phi %
_{1}=0$, and for the traceless EOS, with $T=0$ ($\protect\omega =1$), is
considered. The behavior of $1-b(r)/r$, $10^{2}\protect\rho $ (solid), $%
\protect\rho +p_{r}$ (dotted) and $\protect\rho +p_{t}$ (dashed) versus $%
r/r_{0}$ for $n=5$, are plotted. It is shown that for $\protect\alpha <0$,
the WEC is satisfied near the wormhole throat ($a$) whereas for $\protect%
\alpha >0$ is not ($b$).}
\label{wor2}
\end{figure*}

Now, restricting our discussion to the case of the 5-dimensional space-time,
one finds that 
\begin{equation}
b_{+}(r)=\frac{(-3r^{2}+\sqrt{9r^{4}+4(3r_{0}^{2}+\alpha )\alpha })r}{%
2\alpha }.
\end{equation}%
The EM tensor components for this solution are given by%
\begin{align}
\rho =& \frac{3[(3\kappa -9r^{2})r^{2}-2(3r_{0}^{2}+\alpha )\alpha ]}{\alpha 
{r^{2}}\kappa }, \\
\rho +p_{r}=& \frac{9[5(\kappa -3r^{2})r^{2}-4(3r_{0}^{2}+\alpha )\alpha ]}{%
4\alpha {r^{2}}\kappa }, \\
\rho +p_{t}=& \frac{(45\kappa -135r^{2})r^{2}-28(3r_{0}^{2}+\alpha )\alpha }{%
4\alpha {r^{2}}\kappa },
\end{align}%
respectively, where $\kappa =\sqrt{9r^{4}+4(3r_{0}^{2}+\alpha )\alpha }$.
Imposing different values on $\alpha $, we plot the quantities $1-b(r)/r$, $%
\rho $, $\rho +p_{r}$ and $\rho +p_{t}$ in Fig. \ref{wor2}. Note that the
components of the EM tensor tend to zero as $r$ tends to infinity. Fig. (\ref%
{wor2}-$b$) shows that for $\alpha >0$ the WEC (and also NEC) is violated in
the vicinity of the wormhole throat, but for $\alpha <0$, it can be
satisfied near the wormhole throat as it is shown in Fig. (\ref{wor2}-$a$).
It can also be seen that $\rho $ and $\rho +p_{t}$ have no real root and
therefore are positive everywhere, while $\rho +p_{r}$ possesses a real root
($r_{c}$), where the value $\rho +p_{r}$ is positive in the radial region $%
r_{0}\leq {r}\leq {r_{c}}$. Thus, one may have normal matter in the radial
region $r_{0}\leq {r}\leq {r_{c}}$. 

\begin{figure*}[tbp]
\begin{center}
\subfigure[~$\alpha=-1.6$] {\  \label{k0} 
\includegraphics[width=0.48\textwidth,height=0.3\textheight]{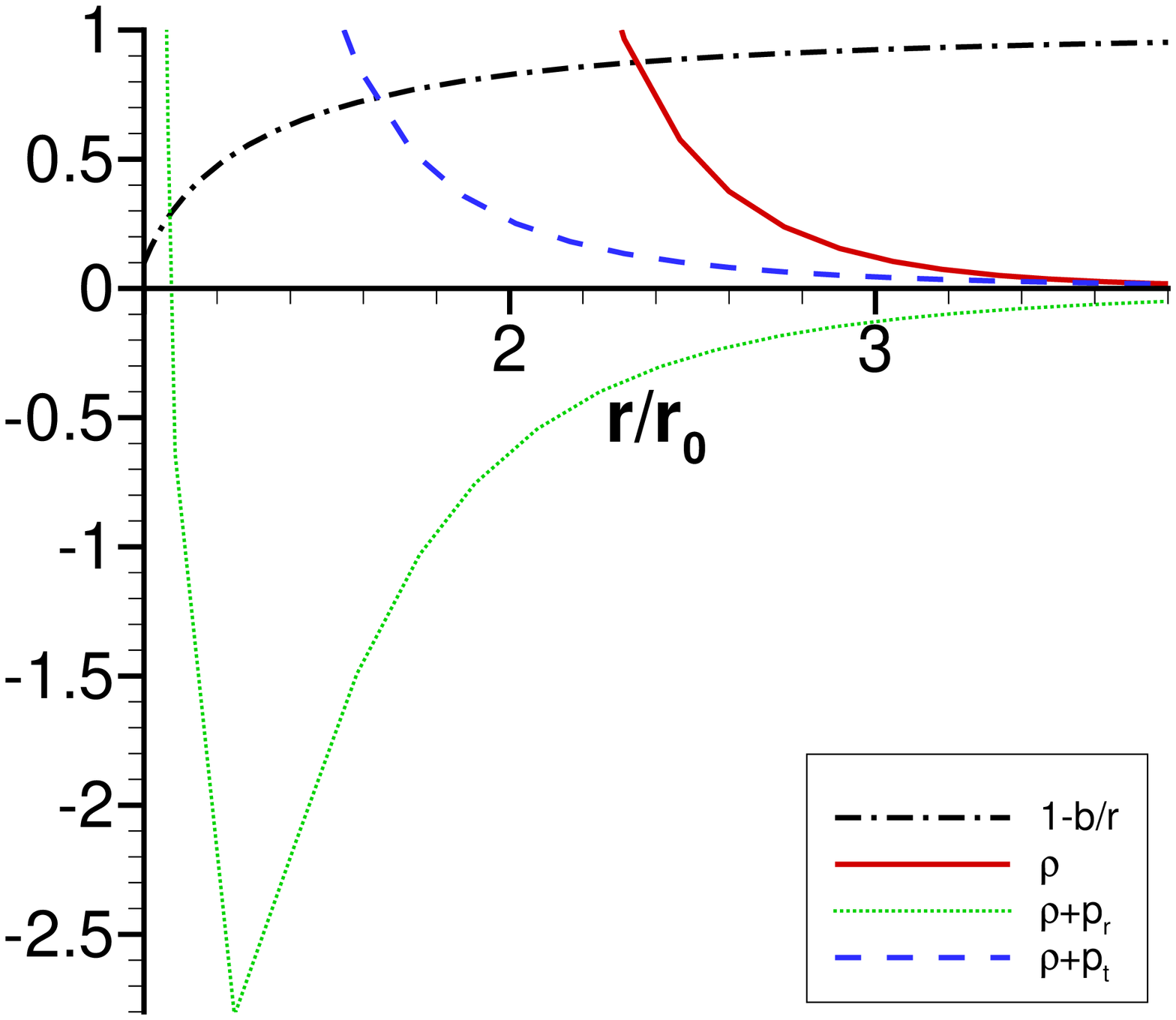}} \text{%
\hspace{0cm}} 
\subfigure[~$\alpha=-3$] {\  \label{k1} 
\includegraphics[width=0.48\textwidth,height=0.3\textheight]{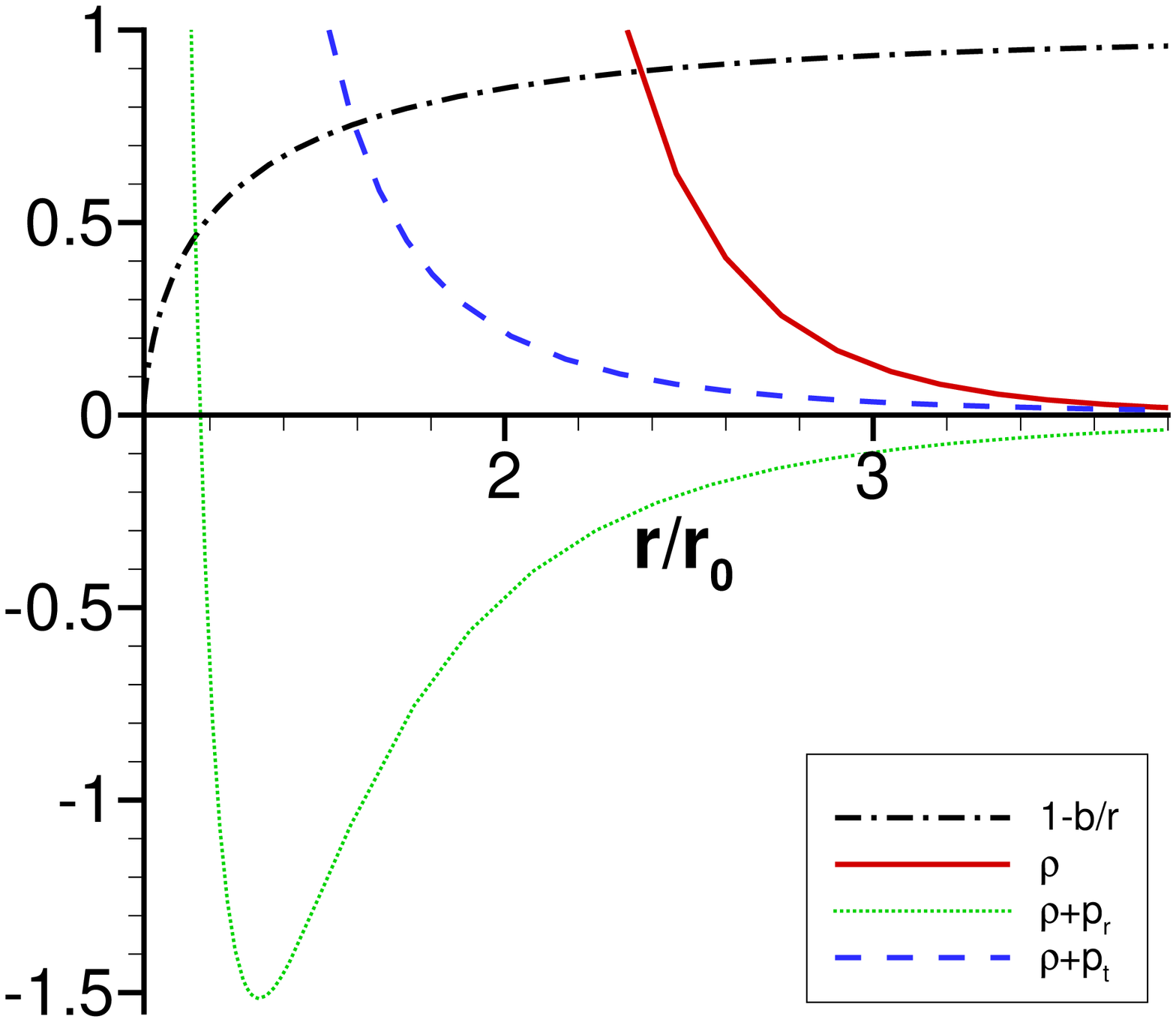}}
\end{center}
\caption{The specific case of a constant redshift function, $\protect\phi %
_{1}=0$, and for the traceless EOS, with $T=0$ ($\protect\omega =1$), is
considered. The behavior of $1-b(r)/r$ (dot-dash), $10^{3}\protect\rho $
(solid), $10(\protect\rho +p_{r})$ (dotted) and $10(\protect\rho +p_{t})$
(dashed) versus $r/r_{0}$ for $n=5$, are plotted. It is shown that the
region of normal matter in the vicinity of the thoroat enlarges as the value
of $\left\vert {\protect\alpha }\right\vert $ increases.}
\label{wor4}
\end{figure*}

Figure \ref{wor4} shows that the increase of $\left\vert {\alpha }%
\right\vert $ enlarges the normal matter region. Briefly, all of these
figures show that it is possible to choose suitable values for the constants
in order to have normal matter in the vicinity of the throat. 

\subsection{Solutions satisfying the WEC}

\label{numsol}

\begin{figure}[tbp]
\includegraphics[width=0.45\textwidth,height=0.3\textheight]{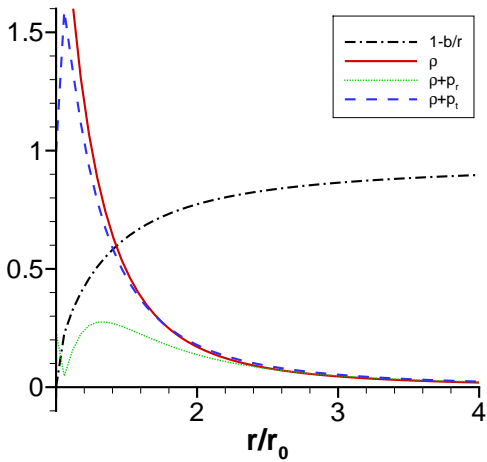}
\caption{The behavior of $1-b(r)/r$ (dot-dash), $\protect\rho $ (solid), $%
\protect\rho +p_{r}$ (dotted) and $\protect\rho +p_{t}$ (dashed) versus $%
r/r_{0}$ for $\protect\alpha =-1.54$ with $\protect\omega =1$ ($T=0$), $%
m=0.3 $, $\protect\phi _{0}=1$, $\protect\phi _{1}=-0.96$, $n=7$ ($b^{\prime
}(r_{0})=-31.59<1$). This solution satisfies the WEC.}
\label{sol1}
\end{figure}

\begin{figure}[tbp]
\includegraphics[width=0.45\textwidth,height=0.3\textheight]{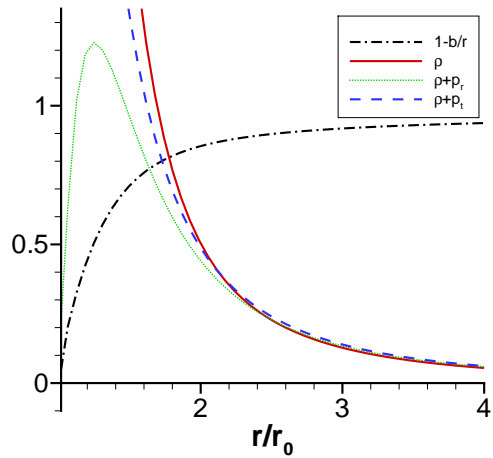}
\caption{The behavior of $1-b(r)/r$ (dot-dash), $\protect\rho $ (solid), $%
\protect\rho +p_{r}$ (dotted) and $\protect\rho +p_{t}$ (dashed) versus $%
r/r_{0}$ for $\protect\alpha =-1.18$ with $\protect\omega =1$ ($T=0$), $%
m=0.28$, $\protect\phi _{0}=1$, $\protect\phi _{1}=-0.99$, $n=11$ ($%
b^{\prime }(r_{0})=-49.35<1$). This solution satisfies the WEC. }
\label{sol2}
\end{figure}

In this section, the equations are solved for the full redshift function of
the form presented in Eq. (\ref{phii}). Since finding exact analytical
solutions is extremely difficult, we consider a simplifying assumption of $%
\omega =1$ so that the EOS reduces to a traceless energy momentum tensor, $%
T=0$. One may now choose specific constant parameters so that the solutions
are asymptotically flat. Although one cannot find explicit solutions for the shape function,
the numerical solutions are plotted in Figs. \ref{sol1} and \ref{sol2}. 
We verify that for these choices, the
quantity $b(r)/r$ tends to zero at spatial infinity, and the behavior of $%
1-b(r)/r$ is plotted in the figures. Note that all of the quantities $\rho
(r)$, $\rho (r)+p_{r}(r)$ and $\rho (r)+p_{t}(r)$ are positive throughout
the spacetime, implying that the WEC is satisfied for all values of $r$.

Note that Bawal and Kar \cite{bha} have claimed that it is not possible to
find wormhole solutions of the GB field equations with normal matter
everywhere. However, this discussion is based on the positivity of the
factor $\left[(b-rb^{\prime })/r+2\phi^{\prime }(b-r)\right]$ [see Eq. (\ref%
{EGBNEC})] throughout spacetime. This factor is indeed positive at the
throat as shown above, but this behaviour is not guaranteed to be positive
for the entire range of the radial coordinate. The solutions presented in
Figs. \ref{sol1} and \ref{sol2} provide a counterexample to the claim in 
\cite{bha}.

Another condition that needs to be satisfied is the condition $b^{\prime
}(r_{0})<1$ at the throat. We verify that the quantity $b^{\prime }_0$ is
given by [see Eq. (\ref{bprim})]: 
\begin{widetext}
\begin{equation}
b^{\prime }(r_{0})=-\frac{2\phi _{0}(n-4)[\alpha (n-7)+r_{0}^{2}(n-2)]+\phi
_{1}[(n-2)(n-4)+m][2\alpha +r_{0}^{2}]}{2\phi _{0}[2\alpha
(n-4)+r_{0}^{2}(n-2)]+\phi _{1}[2\alpha (2(n-2)-m-8)+r_{0}^{2}(2(n-2)-m)]}.
\end{equation}%
\end{widetext}
One can find the values of $b^{\prime }(r_{0})$ for these numerical
solutions in the caption of Figs. \ref{sol1} and \ref{sol2}. 

\section{Summary and Conclusions}

\label{conc}

In this paper, we have explored higher-dimensional asymptotically flat
wormhole solutions in the framework of Gauss-Bonnet (GB) gravity by
considering a specific choice for a radial-dependent redshift function and
by imposing a particular equation of state. We have shown explicitly that
the WEC is satisfied at the throat by considering a negative Gauss-Bonnet
coupling constant, i.e., $\alpha < 0$, and in which the wormhole throat is
constrained by the following condition $r_0^2 < 2|\alpha|$. This confirms
previous results outlined in \cite{bha}. Furthermore, we have briefly
presented solutions in higher-dimensional Einstein gravity, $\alpha=0$, for
a specific radial-dependent redshift function. Furthermore, we have
considered a constant redshift function and shown specifically that, for $%
\alpha<0$, one may have normal matter in a determined radial region $%
r_{0}\leq {r}\leq {r_{c}}$, and that the increase of $\left\vert {\alpha }%
\right\vert $ enlarges the normal matter region.

However, the main motivation of this work resides in finding solutions that
satisfy the WEC throughout the entire spacetime. We have been intrigued by
previous assumptions claiming that wormholes solutions were not possible for
the $\alpha < 0$ case \cite{bha}. However, we agree with the discussion in 
\cite{mae}, in that the nontrivial assumption discussed in \cite{bha}, which
is valid only at the wormhole throat, cannot be extended throughout the
entire range of the radial coordinate. In this work, we provided a
counterexample to this claim, and found for the first time solutions that
satisfy the WEC throughout the entire spacetime. In this context, it would
be interesting to extend the analysis carried out in third order Lovelock
gravity considered in \cite{deh}, where it was shown that a negative third
order coupling constant enlarges the radius of the region of normal matter
relative to the second order theory, and perhaps it may be possible to find
solutions that satisfy the WEC throughout the entire spacetime. Work along
these lines is presently underway.

\acknowledgments{MRM thanks Research
Institute for Astronomy \& Astrophysics of Maragha (RIAAM), Iran, for financial support.
FSNL acknowledges financial  support of the Funda\c{c}\~{a}o para a
Ci\^{e}ncia e Tecnologia through an Investigador FCT Research contract, with
reference IF/00859/2012, funded by FCT/MCTES (Portugal), and the grant
EXPL/FIS-AST/1608/2013. }

\end{document}